\documentclass[manuscript]{aastex6}

\fullcollaborationName{}

\begin{document}

%% LaTeX will automatically break titles if they run longer than
%% one line. However, you may use \\ to force a line break if
%% you desire.

\title{Imaging the photoevaporating disk and radio jet of GM Aur}

%% Use \author, \affil, plus the \and command to format author and affiliation 
%% information.  If done correctly the peer review system will be able to
%% automatically put the author and affiliation information from the manuscript
%% and save the corresponding author the trouble of entering it by hand.
%%
%% The \affil should be used to document primary affiliations and the
%% \altaffil should be used for secondary affiliations, titles, or email.

%% Authors with the same affiliation can be grouped in a single
%% \author and \affil call.
\author{Enrique Mac\'{\i}as\altaffilmark{1},
Guillem Anglada\altaffilmark{1},
Mayra Osorio\altaffilmark{1},
Nuria Calvet\altaffilmark{2},
Jos\'e M. Torrelles\altaffilmark{3, 4},
Jos\'e  F. G\'omez\altaffilmark{1},
Catherine Espaillat\altaffilmark{5},
Susana Lizano\altaffilmark{6},
Luis F. Rodr\'{\i}guez\altaffilmark{6},
Carlos Carrasco-Gonz\'alez\altaffilmark{6},
Luis Zapata\altaffilmark{6}
}

\altaffiltext{1}{Instituto de Astrof\'\i sica de Andaluc\'\i a (CSIC), 
Glorieta de la Astronom\'\i a s/n, E-18008 Granada, Spain;  email: 
{\tt emacias@iaa.es}}
\altaffiltext{2}{Department of Astronomy, University of Michigan,  825 
Dennison Building, 500 Church St, Ann Arbor, MI 48109, USA}
\altaffiltext{3}{Institut de Ci\`encies de l'Espai (CSIC)-Institut de 
Ci\`encies del Cosmos (UB)/IEEC, Mart\'{\i} i Franqu\`es 1, E-08028 
Barcelona, Spain}
\altaffiltext{4}{The ICC (UB) is a CSIC-Associated Unit through the ICE}
\altaffiltext{5}{Department of Astronomy, Boston University, 725 Commonwealth
Avenue, Boston, MA 02215, USA}
\altaffiltext{6}{Instituto de Radioastronom\'{\i}a y Astrof\'{\i}sica 
UNAM, Apartado Postal 3-72 (Xangari), 58089 Morelia, Michoac\'an, Mexico}

%% Use the \and command so offset the last author.
%\and
%% Notice that each of these authors has alternate affiliations, which
%% are identified by the \altaffilmark after each name.  Specify alternate
%% affiliation information with \altaffiltext, with one command per each
%% affiliation.
%\altaffiltext{1}{AAS Journals Data Scientist}

%% Mark off the abstract in the ``abstract'' environment. 
\begin{abstract}
Photoevaporation is probably the main agent for gas dispersal during 
the last stages of protoplanetary disk evolution. However, the overall mass loss 
rate in the photoevaporative wind and its driving mechanism are 
still not well understood. 
Here we report 
multi-configuration Very Large Array observations 
at 0.7, 3, and 5 cm towards the transitional disk of GM Aur. Our radio continuum
observations allow us to image and spatially resolve, for the first time, the 
three main components at work in this stage of the disk evolution: the disk 
of dust, the ionized radio jet perpendicular to it, and the photoevaporative wind 
arising from the disk. The mass loss rate inferred from the flux density of the radio jet 
is consistent with the ratio between ejection and accretion rates 
found in younger objects, suggesting that transitional disks can 
power collimated ejections of material apparently following
the same physical mechanisms as much younger protostars. Our 
results indicate that extreme-UV (EUV) radiation
is the main ionizing 
mechanism of the photoevaporative wind traced by the free-free emission. 
The required low EUV photon luminosity of  $\sim6\times10^{40}$ s$^{-1}$ 
would produce a photoevaporation rate of only $\dot{M}_{\rm w,EUV}\simeq1.3\times10^{-10}~M_{\odot}$ yr$^{-1}$. 
Therefore, other mechanisms are required to disperse the disk in the timescale imposed by observations.
\end{abstract}

%% Keywords should appear after the \end{abstract} command. 
%% See the online documentation for the full list of available subject
%% keywords and the rules for their use.

%% From the front matter, we move on to the body of the paper.
%% Sections are demarcated by \section and \subsection, respectively.
%% Observe the use of the LaTeX \label
%% command after the \subsection to give a symbolic KEY to the
%% subsection for cross-referencing in a \ref command.
%% You can use LaTeX's \ref and \label commands to keep track of
%% cross-references to sections, equations, tables, and figures.
%% That way, if you change the order of any elements, LaTeX will
%% automatically renumber them.

%% We recommend that authors also use the natbib \citep
%% and \citet commands to identify citations.  The citations are
%% tied to the reference list via symbolic KEYs. The KEY corresponds
%% to the KEY in the \bibitem in the reference list below. 

\section{Introduction} \label{sec:intro}

Photoevaporation, together with viscous accretion, is expected to play an important role in the dispersal of protoplanetary disks \citep{wil11,ale14}. High energy radiation -- i.e. far-UV (FUV), extreme-UV (EUV), and X-ray radiation -- originating at the stellar chromosphere of low-mass stars can ionize and heat the disk surface \citep{cla01,gor09,owe10}. Beyond a critical radius, the thermal energy of the heated surface becomes higher than its binding gravitational energy and the gas escapes in the form of a wind. While EUV photons produce a fully-ionized wind, X-rays can penetrate into deeper, neutral regions of the disk, creating a denser, partially-ionized photoevaporative wind \citep{gor09,owe11}.

Although the early stages of disk evolution are dominated by viscous accretion, as the accretion rate decreases, central star-driven photoevaporation should eventually dominate over disk accretion, clearing the gas and leading the disk into the debris disk phase \citep{ale14}. The timescale of gas removal and, thus, the impact of photoevaporation in the disk evolution, will strongly depend on the ionization rate reaching the disk and the mass loss rate produced by the photoevaporative winds.

So far, the primary diagnostic of disk photoevaporation has been optical and mid-IR forbidden line emission (e.g. [O I] $6300$ \AA ~and [Ne II] $12.81~\mu$m) from the wind \citep{fon04}. The redshifted side of the flow is blocked by the disk midplane, which is optically thick at these wavelengths. Therefore, the line profile is expected to be essentially narrow ($\sim10$ km s$^{-1}$) and blueshifted by 5--7 km s$^{-1}$ \citep{fon04}, although high disk inclinations and optically thin regions in the disk (like gaps or cavities) can produce broader lines centered at the systemic velocity \citep{ale08}. Blueshifted lines have been detected in a number of protoplanetary disks\citep{naj09,pas09,sac12}. However, similar line profiles and luminosities can be obtained with different models \citep{ale14}. Therefore, these lines cannot be used to constrain the high energy radiation responsible for the photoevaporative wind or to infer the mass loss rate in the flow \citep{erc10}. 

\citet{lug04}, and more recently \citet{ava12}, proposed that free-free emission at cm wavelengths could be used as a diagnostic of disk photoevaporation in massive stars. \citet{pas12} and \citet{owe13} followed a similar approach focusing on central star-driven photoevaporation in low-mass stars, and proposed that cm observations could actually provide a better observational test than forbidden line observations. The free-free emission from the  fully (EUV case) or partially (X-rays case) ionized disk surface is optically thin and, thus, directly proportional to the ionizing radiation reaching the disk. Since the X-ray luminosity of T Tauri stars can be directly measured, one can in principle estimate the free-free emission produced by the X-ray-ionized gas and, therefore, estimate the EUV photon luminosity impinging on the disk from the remaining observed emission. Following this idea, recent observational studies at cm wavelengths have focused on the free-free emission of protoplanetary disks in order to constrain photoevaporation models \citep{gal14,pas14}. Due to their limited angular resolution, however, it is difficult to ascertain that these radio observations were not contaminated by free-free emission from an accretion-driven collimated jet. Since classical T Tauri stars present lower accretion rates than younger stellar objects, weak (or even absent) radio jet emission is expected in this type of sources. Nevertheless, thanks to the improved sensitivity of the Karl G. Jansky Very Large Array (VLA), \citet{rod14} were recently able to resolve the emission at 3.3 cm of a relatively weak radio jet in AB Aur, a Herbig Ae star surrounded by a transitional disk.

GM Auriga is a well-known T Tauri star ($d\simeq$140 pc, K5 spectral type, $L_{\star}\simeq0.9$ $L_{\odot}$, $M_{\star}\simeq1.1~M_{\odot}$; \citealp{ken95}) surrounded by a transitional disk with a dust cavity of radius $\sim24$ au ($\sim0.17''$; \citealp{cal05,hug09,esp10}). [O I] and [Ne II] lines have been detected towards GM Aur, indicating the presence of high energy radiation reaching the disk. However, the [OI] spectrum has a very poor spectral resolution and the [NeII] spectrum shows no clear evidence of a blueshifted line peak that would confirm the presence of photoevaporation in GM Aur \citep{har95,naj09}. This could indicate that the lines are tracing a bounded ionized layer of the disk \citep{naj09}. Alternatively, the lack of an observed blueshifted line peak could actually be due to the disk cavity, which could allow the redshifted component of the wind to be visible \citep{owe13}, or due to an insufficient signal-to-noise ratio in the [Ne II] spectrum. Therefore, even though observations indicate that high energy radiation is impinging on the disk surface, the presence of photoevaporation in GM Aur is still uncertain.

Here we report new sensitive high angular resolution VLA observations at 7 mm, 3 cm, and 5 cm towards the transitional disk of GM Aur, showing evidence of the presence of free-free emission from both photoevaporative winds and a radio jet.

\section{Observations} \label{sec:observations}

Our new observations were carried out with the VLA of the National Radio Astronomy Observatory (NRAO)\footnote{The NRAO is a facility of the National Science Foundation operated under cooperative agreement by Associated Universities, Inc.} using three different configurations (A, B, and C) at Q (7 mm), X (3 cm), and C (5 cm) bands. We also reduced VLA archival data at Ka, K, and C bands (see Table \ref{Tab:obs}). For all the observations, 3C147 and J0438+3004 were used as the amplitude and complex gain calibrators, respectively. 3C84 was observed at Q band in order to calibrate the bandpass and delays. For the rest of the bands 3C147 was used as the bandpass calibrator. The expected uncertainty in the absolute flux calibration is $\sim10\%$.

%\floattable
\begin{deluxetable*}{lccccccc}
\tablecaption{VLA observations \label{Tab:obs}}
\tablehead{
\colhead{Project} & \colhead{Observing} &\colhead{} & \colhead{} & \colhead{Central} & \colhead{Bandwidth} & \colhead{On-source}\\
\colhead{Code} & \colhead{Date} & \colhead{Conf.} & \colhead{Band} & \colhead{Frequency} & \colhead{} & \colhead{time}\\
\colhead{} & \colhead{} & \colhead{} & \colhead{} & \colhead{(GHz)} & \colhead{(GHz)} & \colhead{(min)}\\
}
\startdata
%-----------------------------------------------------------------------------------------
15B-352 & 2015-Sep-28 &  A    &      C    &                6          &         4         & 39.6  \\
        & 2015-Sep-10 &       &           &                           &                   & 39.6  \\
15B-352 & 2015-Sep-26 &  A    &      X    &               10          &         4         & 38.6  \\
        & 2015-Sep-11 &       &           &                           &                   & 38.6  \\
14B-285 & 2015-May-09 &  B    &      X    &               10          &         4         & 41.8  \\
14B-285 & 2014-Oct-16 &  C    &      Q    &              44           &         8         & 24.2  \\
14B-285 & 2014-Oct-16 &  C    &      X    &               10          &         4         & 8.6   \\
AC982   & 2011-Jul-25 &  A    &      C    &               6           &         2         & 22.3  \\
BL175   & 2011-May-22 &  BnA  &      C    &               6           &         2         & 3.6   \\
BL175   & 2011-Apr-25 &  B    &      C    &               6           &         2         & 3.5   \\
        & 2011-Mar-06 &       &           &                           &                   & 3.6   \\
AC982   & 2010-Nov-14 &  C    &     Ka    &          30.5, 37.5       &       2$\times$2  & 94.5  \\
AC982   & 2010-Sep-13 &  D    &     Ka    &             32.7          &         2         & 3.0   \\
        & 2010-Sep-10 &       &           &                           &                   & 2.3   \\
AC982   & 2010-Sep-11 &  D    &      K    &              21           &         2         & 24.0   \\
        & 2010-Aug-24 &       &           &                           &                   & 4.8  \\
%-----------------------------------------------------------------------------------------
\enddata
\tablecomments{Observations from projects AC982 and BL175 were taken from the VLA archive.}
\end{deluxetable*}

Data calibration was performed using the VLA pipeline integrated in the data reduction package Common Astronomy Software Applications (CASA; version 4.3.1)\footnote{https://science.nrao.edu/facilities/vla/data-processing}. After running the pipeline, we inspected the calibrated data, performed additional data flagging and re-ran the pipeline as many times as needed. Cleaned images were obtained with the CLEAN task of CASA by using the multi-scale multi-frequency deconvolution algorithm described in \citet{rau11}. In order to get a better frequency coverage we also split in frequency the observations at Q and X bands, obtaining images at 6.5 mm, 7.1 mm, 2.7 cm, and 3.3 cm. 

\section{Results} \label{sec:results}

\begin{figure}
\figurenum{1}
\plotone{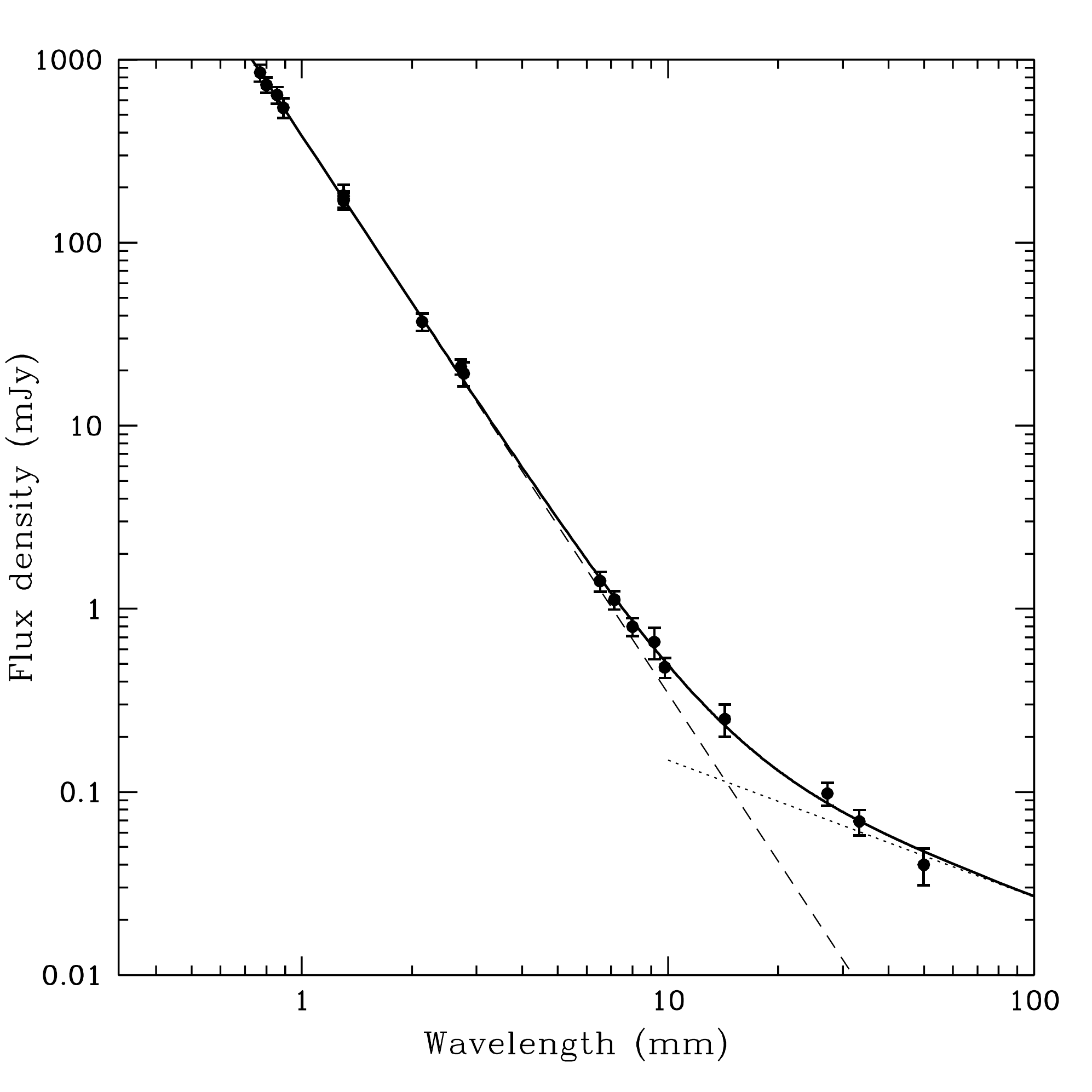}
\caption{Spectral energy distribution (SED) of the radio emission of GM Aur. The data points and error bars are listed in Table \ref{Tab:fluxes}. The lines indicate the fit to the SED (dashed: dust thermal emission; dotted: free free emission; solid: total).\label{fig:sed}}
\end{figure}

We have compiled the flux densities ($S_{\nu}$) of GM Aur at different (sub-)mm and cm wavelengths, both from our new data and from the literature (Table \ref{Tab:fluxes}), and have constructed the spectral energy distribution (SED) shown in Fig. \ref{fig:sed}. As can be seen, the emission at cm wavelengths is above the expected contribution from the dust. The flat spectral index ($\alpha<2$, where $S_{\nu}\propto\nu^{\alpha}$) between 5 and 2 cm cannot be explained with only dust emission, even assuming large (cm-sized) dust grains. By fitting the sum of two power laws to the SED from 0.89 mm to 5 cm, the emission can be explained as a combination of dust thermal ($\alpha_d=3.05\pm0.14$) and free-free emission from ionized gas ($\alpha_{\rm ff}=0.75\pm0.13$). The positive spectral index of the spectrum between 5 and 3 cm, as well as the fact that the source at 3 cm appears to be extended, indicate that the contribution of possible non-thermal emission is negligible.

%\floattable
\begin{deluxetable}{cccc}
\tablecaption{GM Aur Flux Densities \label{Tab:fluxes}}
\tablehead{
\colhead{Frequency} & \colhead{Wavelength} & \colhead{Flux\tablenotemark{a}} & \colhead{}\\
\colhead{(GHz)} & \colhead{(mm)} & \colhead{(mJy)} & \colhead{Ref.} \\
}
\startdata
%-----------------------------------------------------------------------------------------
480             &            0.623        &       1300$\pm$300         &     2 \\
390             &            0.769        &       850 $\pm$ 90         &     2 \\
375             &            0.800        &       730 $\pm$ 70         &     1 \\
350             &            0.856        &       640 $\pm$ 70         &     6 \\
337             &            0.89         &       550 $\pm$ 70         &     7 \\
230             &            1.3          &       170 $\pm$ 17         &     3 \\
230             &            1.3          &       180 $\pm$ 30         &     6 \\
230             &            1.3          &       173 $\pm$ 19         &     7 \\
141             &            2.13         &        37 $\pm$ 4          &     5 \\
110             &            2.72         &        21 $\pm$ 3          &     6 \\
108             &            2.77         &        19 $\pm$ 3          &     4 \\
46.0            &            6.52         &1.42 $\pm$ 0.18\tablenotemark{b}& 8 \\
42.0            &            7.14         &1.12 $\pm$ 0.13\tablenotemark{b}& 8 \\
37.5            &            8.0          &       0.80 $\pm$ 0.09      &     8 \\
32.7            &            9.17         &       0.66 $\pm$ 0.13      &     8 \\
30.5            &            9.8          &       0.48 $\pm$ 0.06      &     8 \\
21.0            &           14.3          &       0.25 $\pm$ 0.05      &     8 \\
11.0            &           27.25         &       0.098 $\pm$ 0.014    &     8 \\
9.0             &           33.3          &       0.069 $\pm$ 0.011    &     8 \\
6.0             &           50.0          &       0.040 $\pm$ 0.009    &     8 \\
%-----------------------------------------------------------------------------------------
\enddata
\tablenotetext{a}{~The uncertainties include the absolute flux calibration uncertainty.}
\tablenotetext{b}{~Previous measurements of the flux density at 7 mm, consistent within the uncertainties with our measurements, were obtained by \citet{rod06}. We adopted the values obtained from our observations because they have a higher sensitivity.}
\tablerefs{(1) \citet{wei89}; (2) \citet{bec91}; (3) \citet{dut98}; (4) \citet{loo00}; (5) \citet{kit02}; (6) \citet{hug09}; (7) \citet{and13}; (8) This paper. }
\end{deluxetable}

\begin{figure}
\figurenum{2}
\gridline{\fig{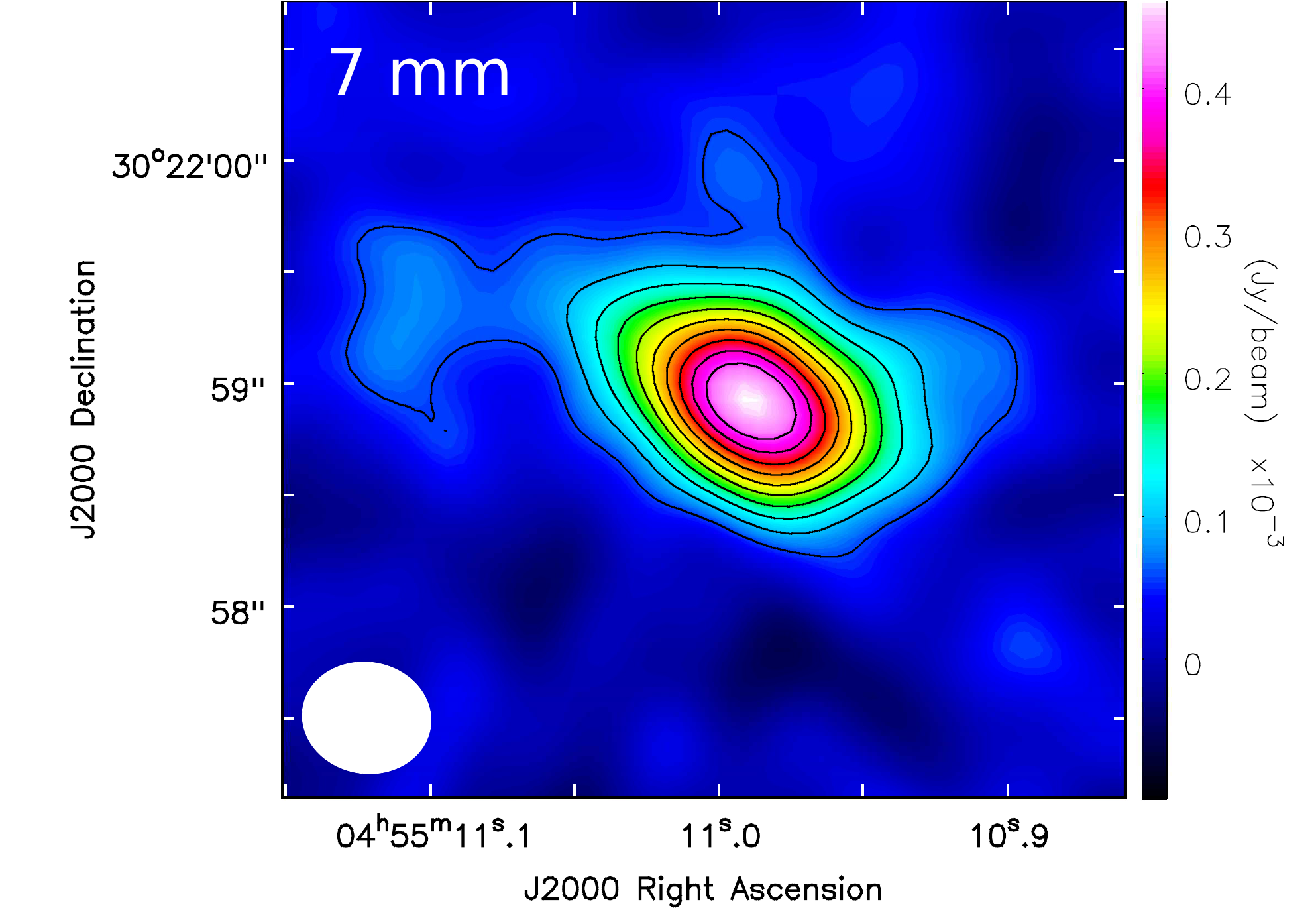}{0.5\textwidth}{\textbf{(a)}}
          \fig{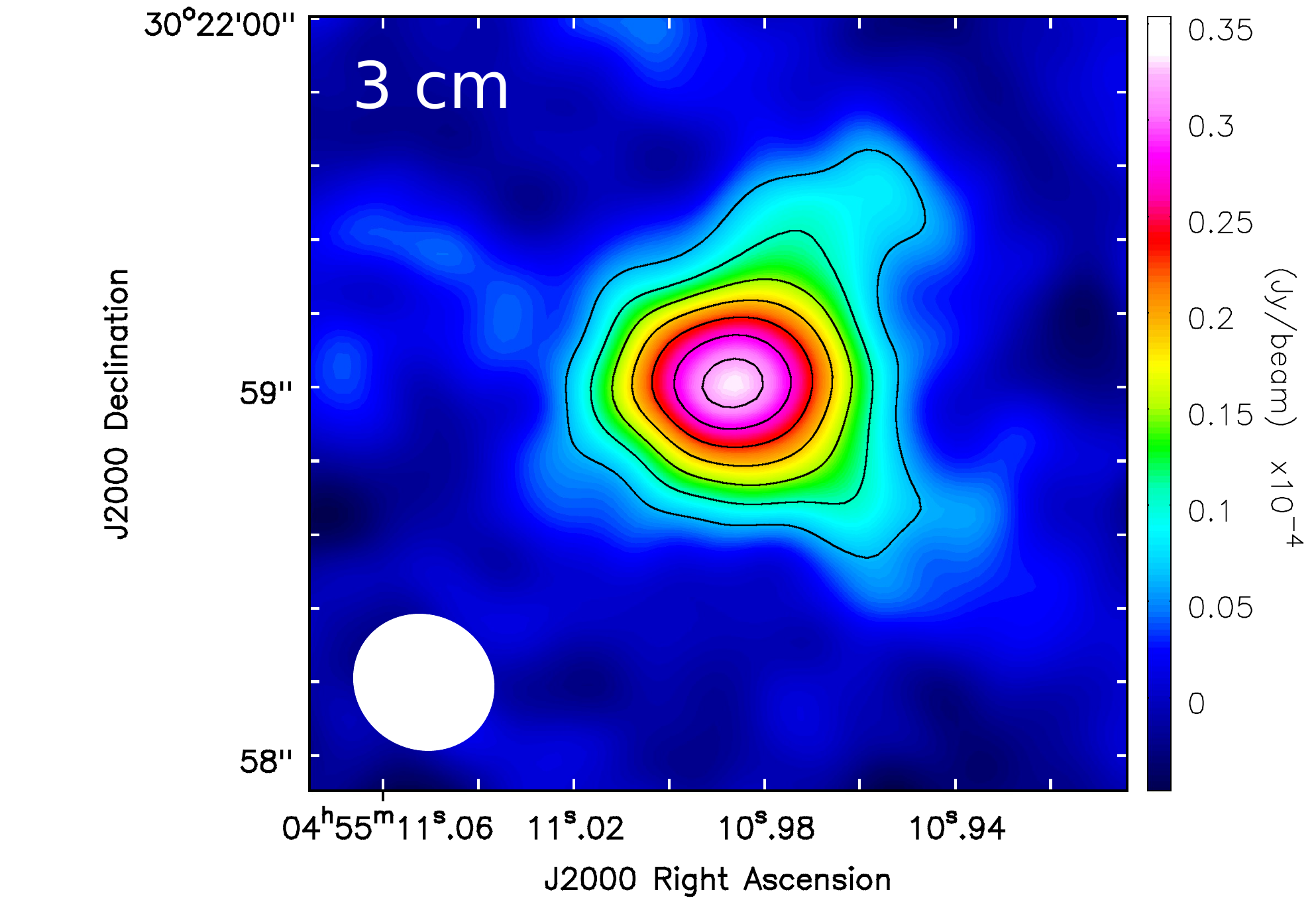}{0.5\textwidth}{\textbf{(b)}}}
\label{fig:maps}
\caption{\textbf{(a)} Naturally weighted VLA image at 7 mm of the transitional disk around GM Aur (synthesized beam=$0\rlap.''58\times0\rlap.''50$, PA=$81^{\circ}$; shown in the lower-left corner). Contour levels are $-$3, 3, 5, 7, 9, 11, 13, 15, 17, 19, and 21 times the rms of the map, $19.3~\mu$Jy beam$^{-1}$. \textbf{(b)} Naturally weighted VLA image of the 3.0 cm emission obtained by combining data from A and B configurations (synthesized beam=$0\rlap.''35\times0\rlap.''32$, PA=$56^{\circ}$; shown in the lower-left corner). Contour levels are $-$3, 3, 5, 7, 9, 11, 13 and 15 times the rms of the map, $2.1~\mu$Jy beam$^{-1}$.}
\end{figure}

The measured flux density at 7 mm, obtained using the full bandwidth of 8 GHz, is $\sim1.25\pm0.14$ mJy, consistent with previous VLA observations \citep{rod06}. From our SED fit at cm wavelengths, we expect a free-free contribution at 7 mm of $\sim 0.2$ mJy, which represents only a $\sim 16 \%$ of the emission at this wavelength. Thus, at 7 mm we are mainly detecting the dust emission. Our 7 mm map shows a resolved disk structure (Fig. \ref{fig:maps}a), although the angular resolution is not enough to reveal the inner cavity of radius $\sim24$ au \citep{cal05,hug09}. From a Gaussian fit to the 7 mm image, we obtained a deconvolved disk size of $\sim220$ au$~\times110$ au (measured at the 3$\sigma$ level), with its major axis oriented in the NE-SW direction (PA$\simeq60^{\circ}$), similar to previous studies at (sub-)mm wavelengths \citep{rod06,hug09}. 

\begin{figure}
\figurenum{3}
\plotone{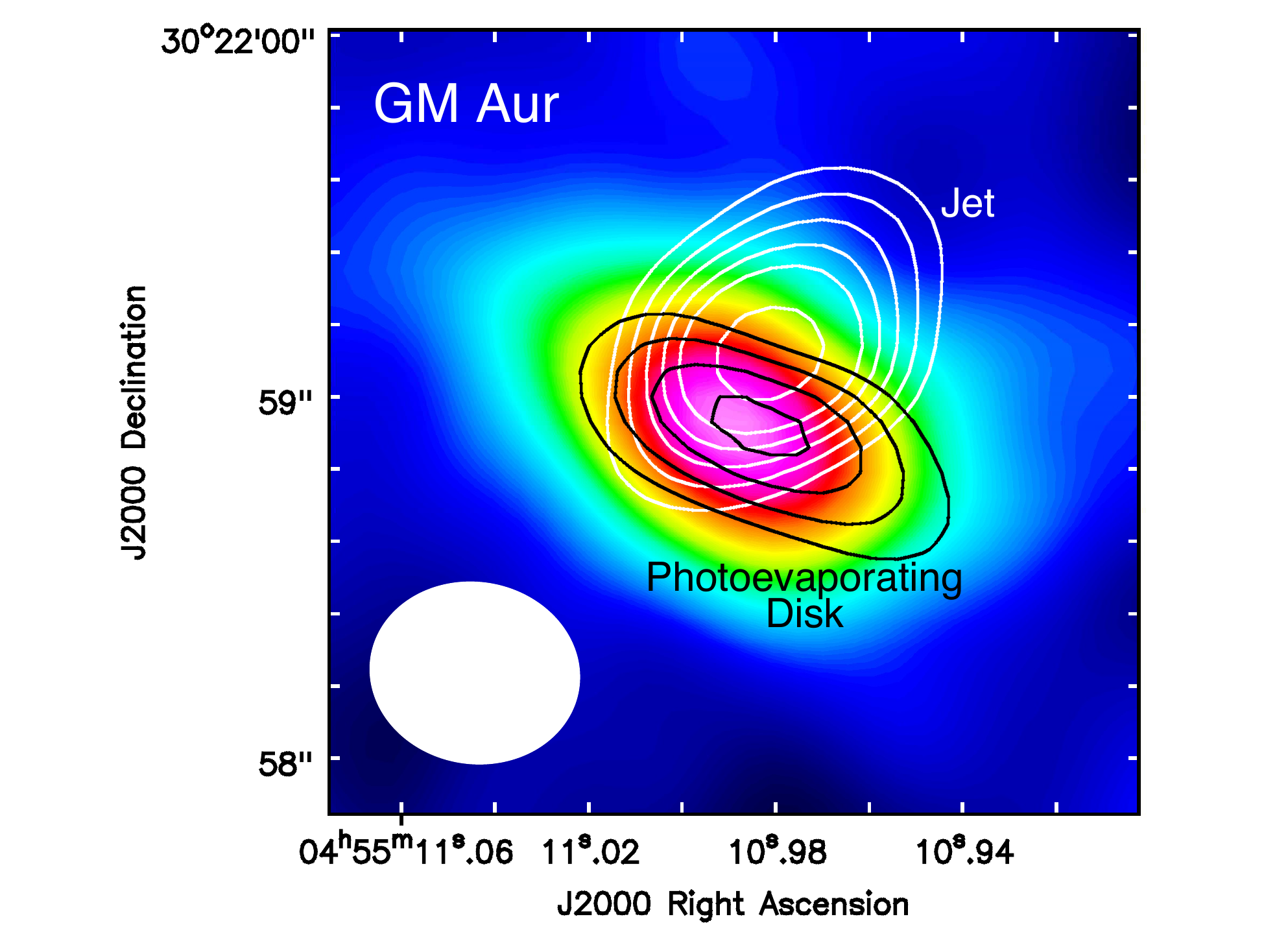}
\label{fig:mapsfit}
\caption{Superposition of the two components of the free-free emission at 3.0 cm obtained from our two-Gaussian fit (black and white contours) over the 7 mm emission of the transitional disk of GM Aur shown in Fig. \ref{fig:maps}a (color scale). We propose that the disk component (black contours) is produced by the photoevaporative winds arising from the disk, while the perpendicular component (white contours) is tracing the free-free emission from an accretion driven radio jet. The 3.0 cm image was previously convolved to the same angular resolution ($0\rlap.''58\times0\rlap.''50$, PA=$81^{\circ}$; shown in the lower-left corner) as the 7 mm image. Contour levels are $-$3, 3, 4, 5, 6, 7, and 9 times the rms of the convolved 3.0 cm image, $2.8~\mu$Jy beam$^{-1}$.}
\end{figure}

An image of the 3.0 cm emission is presented in Figure \ref{fig:maps}b. The emission shows a tripolar structure, suggesting a combination of dust emission from a disk and free-free emission from a monopolar (one-sided) radio jet perpendicular to it. This kind of morphology, with two perpendicular structures, has already been observed in other sources, such as HH 111 \citep{rod08} and HH 80-81 \citep{car12}. In both cases, the emission was separated into two components by fitting two Gaussians to it. In the case of GM Aur, we estimate from the SED fit a dust contribution at 3.0 cm of $\sim11~\mu$Jy, which represents only $1/8$ of the detected emission at 3.0 cm. In order to check whether this flux density can account for the disk emission of GM Aur at 3.0 cm, we obtained an image of the estimated dust emission at 3.0 cm by scaling the 7 mm image with the dust spectral index obtained from the SED fit. We then subtracted this scaled image to the 3.0 cm image, which was previously convolved to have the same angular resolution as the 7 mm observations. The resulting image still showed significant emission from the disk. By fitting two Gaussians to the remaining 3.0 cm emission, we separated the source into a component with the same PA as the GM Aur disk ($S_{\nu}=31\pm3~\mu$Jy), and another one perpendicular to it ($S_{\nu}=45\pm3~\mu$Jy)\footnote{We note that none of the fit parameters were fixed and yet two perpendicular components were obtained.} (Fig. \ref{fig:mapsfit}). We interpret these two components to be tracing the free-free emission from a photoevaporating disk and a radio jet, respectively (see \S4). 

We also combined our new 5 cm observations with archival data (Table \ref{Tab:obs}) to get a  sensitive image (rms$\simeq2.5~\mu$Jy; beam=$0\rlap.''53\times0\rlap.''36$, PA=$-74^{\circ}$). The image shows a compact source at the position of the star, although its size is sensitivity limited. Hence, we were not able to constrain the size of the source at this wavelength. The rest of the archival observations, at Ka and K bands, were obtained with relatively low angular resolution, so they showed an unresolved source at the position of GM Aur.

\section{Discussion} \label{sec:discussion}

The excess of free-free emission at cm wavelengths in GM Aur has been previously attributed, as well as in other transitional disks, to photoevaporative winds arising from the disk \citep{pas12,owe13}. Our 3.0 cm observations spatially resolve the emission, indicating the presence of two different components of free-free emission, one with the same PA as the GM Aur disk and another one perpendicular to it. The morphology of the disk component, as well as its flat spectral index between 5 and 2 cm, indeed suggests the presence of photoevaporative winds in GM Aur. However, our results indicate that a similar fraction of the free-free emission is arising from the perpendicular component, whose morphology, position and spectral index suggest that it is tracing an accretion-driven radio jet. In the following we discuss in more detail the nature of both free-free components.

\subsection{Accretion-driven radio jet}

As mentioned above, our 3.0 cm observations indicate the presence of a radio jet in GM Aur. A similar result was recently found in AB Aur \citep{rod14}. Even though AB Aur is associated with a transitional disk, it has a relatively high mass accretion rate of $\dot{M}_{{\rm acc}}\simeq1.4\times10^{-7}~M_{\odot}$ yr$^{-1}$ \citep{sal13}. GM Aur, on the other hand, presents a much lower $\dot{M}_{{\rm acc}}\simeq(0.4$-$1)\times10^{-8}~M_{\odot}$ yr$^{-1}$ \citep{ing15}. Despite this difference of more than one order of magnitude, our high sensitivity observations have allowed us to detect a radio jet in GM Aur. Therefore, our results indicate that transitional disks, even those with very low mass accretion rates, may be associated with ionized jets.

From our two-Gaussian fit we estimated a flux density of $\sim45~\mu$Jy at 3.0 cm for the radio jet of GM Aur. Following a similar analysis to that of \citet{rod14}, we can test whether this jet follows the empirical correlation obtained by \citet{ang95} and \citet{ang15} between the radio luminosity ($S_{\nu}d^2$) at 3.6 cm and the bolometric luminosity ($L_{bol}$). Using $L_{bol}\simeq0.9~L_{\odot}$ for GM Aur as well as the free-free spectral index obtained in \S3, the correlation predicts a flux density at 3.0 cm of $\sim440~\mu$Jy, which is a factor of $\sim10$ higher than our measured value. A similar result was found by \citet{rod14} in AB Aur, where the flux density obtained with the correlation was a factor $\sim20$ higher than the measured value. The reason for this discrepancy is attributed to the fact that the correlation was obtained for younger sources (class 0 and I), where the bolometric luminosity is dominated by the accretion luminosity. On the contrary, in sources with transitional disks, which have much lower mass accretion rates, the luminosity is dominated by the stellar component. These results suggest that the previous correlation actually relates the accretion luminosity, traced by $L_{bol}$ in very young objects, and the ejection of material, traced by the free-free emission of the radio jet.

The accretion and outflow rates can be compared applying the empirical correlation between the momentum rate of the outflow ($\dot{P}_{\rm out}$) and the radio luminosity \citep{ang95,ang15}:
\begin{equation}
\left(\frac{S_{\nu}d^2}{{\rm mJy~kpc}^2}\right)=190\left(\frac{\dot{P}_{\rm out}}{M_{\odot}{\rm ~yr}^{-1}{\rm~km~s}^{-1}}\right)^{0.9}.
\end{equation}
Using our estimated flux density of $\sim45~\mu$Jy for the radio jet at 3.0 cm we obtain $\dot{P}_{{\rm out}}\simeq1.0\times10^{-6}~M_{\odot}$ yr$^{-1} $ km s$^{-1}$. Assuming a jet terminal velocity of $\sim200$--$300$ km s$^{-1}$ \citep{cab07}, we estimate a mass loss rate $\dot{M}_{{\rm out}}\simeq3$--$5\times10^{-9}~M_{\odot}$ yr$^{-1}$ for the jet of GM Aur, one of the lowest mass loss rates ever obtained for a jet. \citet{ing15} measured a variable mass accretion rate in GM Aur of $\dot{M}_{{\rm acc}}=0.4$--$1.1\times10^{-8}~M_{\odot}$ yr$^{-1}$ within a time span of $\sim3.5$ months. Taking into account this variability in the accretion rate, as well as the observed dispersion in the empirical correlation derived by \citet{ang95} and \citet{ang15}, we conclude that our estimated mass loss rate for the jet of GM Aur is consistent with a ratio $\dot{M}_{\rm out}/\dot{M}_{\rm acc}\simeq0.1$, typical of younger objects \citep{cab07}. A similar result was obtained by \citet{rod14} for the transitional disk of AB Aur. Thus, our results indicate that, even though accretion decreases as a star evolves, the ratio between accretion rate and outflow rate in the jet remains similar, suggesting that a similar ejection mechanism persists during the whole process of star formation.

\subsection{Photoevaporating Disk}

\citet{pas12} studied the free-free emission at cm wavelengths produced by a photoevaporating disk assuming that the disk is heated by EUV or X-ray radiation. They obtained the following relations:
\begin{equation}
\left(\frac{S_{\nu}}{\mu \rm Jy}\right)=2.9\times10^{-39}\left(\frac{51~\rm pc}{d}\right)^2\left(\frac{\Phi_{\rm EUV}}{\rm s^{-1}}\right),
\end{equation}
\begin{equation}
\left(\frac{S_{\nu}}{\mu \rm Jy}\right)=2.4\times10^{-29}\left(\frac{51~\rm pc}{d}\right)^2 \left(\frac{L_{X}}{\rm erg~s^{-1}}\right).
\end{equation}
where $S_{\nu}$ is the flux density at 3.5 cm, $\Phi_{\rm EUV}$ is the EUV photon luminosity, and $L_{X}$ is the X-ray luminosity \citep{pas12}. \citet{owe13} extended this analysis with numerical calculations and obtained similar results. From equation (3), one can estimate the free-free emission produced by the X-ray-heated gas if the stellar X-ray luminosity ($L_X$) is known. The measured $L_X$ for GM Aur is $\sim1.6\times10^{30}$ erg s$^{-1}$ \citep{gud10}, which would produce a flux density of only $\sim5.7~\mu$Jy at 3.0 cm. This value is much lower than our estimated flux density of $\sim31\pm3~\mu$Jy at 3.0 cm for the free-free emission of the photoevaporating disk, which indicates that in GM Aur the photoionization of the gas cannot be only produced by X-rays, and should be mainly due to EUV radiation. Following equation (2), we obtain that a $\Phi_{\rm EUV}\simeq5.8\times10^{40}$ s$^{-1}$ is required to account for the remaining $\sim25~\mu$Jy of the observed free-free emission from the photoevaporative wind in GM Aur. 

Since EUV photons are highly absorbed by H, only indirect measurements of $\Phi_{\rm EUV}$ from T Tauri stars can be performed. Using forbidden line observations, EUV photon luminosities $\sim10^{41}$-$10^{44}$ s$^{-1}$ have been inferred \citep{ale05,esp13}. Recently, following \citet{pas12} prescriptions, \citet{gal14} and \citet{pas14} estimated $2\times10^{40}\lesssim\Phi_{\rm EUV}\lesssim10^{42}$ s$^{-1}$ using radio observations of different transitional disks. Our estimated EUV photon luminosity for GM Aur is close to the lower limit of these latter measurements. However, none of the previous studies took into account the possible contribution of a radio jet. Our observations have allowed us to separate both components of the free-free emission, resulting in a lower but probably more accurate estimate of $\Phi_{\rm EUV}$. 

In particular, \citet{owe13} and \citet{pas14} estimated a higher value of $\Phi_{\rm EUV}\simeq2\times10^{41}$ s$^{-1}$ for GM Aur based mostly on Arcminute Microkelvin Imager Large Array (AMI-LA) observations. Due to the low angular resolution of their observations (beam=$39\rlap.''4 \times25\rlap.''0$), these authors were not able to separate the radio jet emission reported in this paper or the contribution of nearby sources that could fall within the large beam of AMI-LA from the free-free emission from the photoevaporative winds. Our 3.0 cm observations show, in fact, a nearby source ($\sim28''$ away) with almost the same flux density as GM Aur that was not separated by the AMI-LA observations. 

It is worth noting that the $\Phi_{\rm EUV}$ estimated from the observed photoionized gas actually represents the EUV radiation impinging on the disk, which could differ from the EUV radiation produced by the star. \citet{hol09} showed that accretion flows and jets could significantly shield the disk from EUV and soft ($\sim0.1$ keV) X-ray radiation. According to their model, EUV and soft X-ray photons can completely cross a jet only if its mass loss rate is $\dot{M}_{out}\lesssim8\times10^{-10}~M_{\odot}$ yr$^{-1}$. In the case of GM Aur, our estimated $\dot{M}_{\rm out}\simeq(3$-$5)\times10^{-9}~M_{\odot}$ yr$^{-1}$in the jet is above this value. This implies that the $\Phi_{\rm EUV}$ reaching the disk and, therefore, our estimated values, could be significantly lower than the $\Phi_{\rm EUV}$ produced by the star. As the accretion rate and, thus, the ejection rate of the jet decrease, the high-energy radiation impinging on the disk could increase, accelerating the photoevaporation of the disk. A larger sample of protoplanetary disks with a good determination of both $\Phi_{\rm EUV}$ and $\dot{M}_{\rm out}$ of the jet would be needed to test this scenario and its impact on the timescale of disk dispersal.

In any case, the low $\Phi_{\rm EUV}$ estimated from radio observations indicates that EUV radiation alone is probably not enough to disperse the disk in the timescale required by observations \citep{pas14}. Equation (4) in \citet{ale14} can be used to estimate the mass loss rate in a fully-ionized photoevaporative wind launched purely by EUV radiation:
\begin{equation}
\left(\frac{\dot{M}_{\rm w,EUV}}{M_{\odot}~\rm yr^{-1}}\right)\simeq1.6\times10^{-10} \left(\frac{\Phi_{\rm EUV}}{10^{41}~\rm s^{-1}}\right)^{1/2} \left(\frac{M_{\ast}}{M_{\odot}}\right)^{1/2}.
\end{equation}
From this equation we obtain that the estimated $\Phi_{\rm EUV}$ for GM Aur would launch a photoevaporative wind with a mass loss rate of only $\dot{M}_{\rm w,EUV}\simeq1.3\times10^{-10}~M_{\odot}$ yr$^{-1}$. This value is much lower than the measured mass accretion rate of $\dot{M}_{{\rm acc}}=(0.4$--$1.1)\times10^{-8}~M_{\odot}$ yr$^{-1}$ \citep{ing15}, suggesting that EUV photons are currently not able to release a significant amount of gas from the disk. Alternatively, X-rays can penetrate much deeper into the disk than EUV radiation and, thus, they can launch a denser, but only partially ionized, photoevaporative wind \citep{gor09,owe11}. A quantitative estimate of the mass loss rate from X-ray photoevaporation ($\dot{M}_{w,X}$) is still subject to significant uncertainties, since several input parameters and discrepancies between the models are not yet well understood \citep{ale14}. Nonetheless, theoretical models predict mass loss rates between 1 and 2 orders of magnitude higher than those produced by EUV radiation for a star like GM Aur. Therefore, even though our results indicate that X-rays are not contributing significantly to the ionization of the photoevaporative wind, according to the models they could be responsible for most of the mass loss rate of the photoevaporating disk \citep{pas14}.
 
\section{Summary and conclusions} \label{sec:conclusions}

We have analyzed the results of multi-configuration VLA observations at Q, Ka, K, X, and C bands towards the transitional disk of GM Aur, revealing the presence of dust thermal and free-free emission at cm wavelengths. 

At 3 cm the emission presents an angularly resolved tripolar morphology that we separate into three components: the dust emission from the GM Aur disk, the free-free emission from a radio jet perpendicular to it, and the free-free emission from a photoevaporative wind arising from the disk. This is the first time that free-free emission from disk photoevaporation in a low mass star has been spatially resolved and separated from other components. 

We conclude that extreme-UV (EUV) 
radiation is the main agent responsible for the ionization of the photoevaporative wind in GM Aur, although requiring a low photon luminosity of $\Phi_{\rm EUV}\simeq6\times10^{40}$ s$^{-1}$. This low EUV photon luminosity produces a mass loss rate of only $\dot{M}_{w,EUV}\simeq1.3\times10^{-10}~M_{\odot}$ yr$^{-1}$. Therefore, other mechanisms, such as X-ray photoevaporation, are required to disperse the disk in the timescale imposed by observations.

On the other hand, we estimate a mass loss rate in the radio jet in GM Aur of $\dot{M}_{{\rm out}}\simeq(3$-$5)\times10^{-9}~M_{\odot}$ yr$^{-1}$, which represents one of the lowest mass ejection rates in a jet estimated so far. Nevertheless, the ratio $\dot{M}_{out}/\dot{M}_{acc}\simeq0.1$, typical of younger protostars, seems to be valid as well for GM Aur. Therefore, our results suggest that disks with very low mass accretion rates still present collimated ejections of material, apparently following the same physical mechanisms as much younger protostars
 
At least in GM Aur, the cm free-free emission of the jet and the photoevaporative wind seem to be of the same order. Future radio observations aiming to study photoevaporation in the last stages of disk evolution should be cautious and try to disentangle the contribution to the observed radio emission of the dust, the jet, and the photoevaporative winds.

%% If you wish to include an acknowledgments section in your paper,
%% separate it off from the body of the text using the \acknowledgments
%% command.
\acknowledgments

We thank the referee, Richard Alexander, for his useful and clarifying comments that improved the paper. E.M., G.A., M.O., J.M.T, and J.F.G. acknowledge support from MINECO (Spain) grant AYA2014-57369-C3 (co-funded with FEDER funds). S.L. acknowledges support from DGAPA-UNAM IN105815 and CONACyT 238631. C.C.-G. acknowledges support from UNAM-DGAPA PAPIIT IA102816. L.Z. acknowledges the financial support from DGAPA, UNAM, and CONACyT (Mexico).

%% To help institutions obtain information on the effectiveness of their 
%% telescopes the AAS Journals has created a group of keywords for telescope 
%% facilities. 

%% Following the acknowledgments section, use the following syntax and the
%% \facility{} macro to list the keywords of facilities used in the research 
%% for the paper.  Each keyword is check against the master list during
%% copy editing.  Individual instruments can be provided in parentheses,
%% after the keyword, but they are not verified.

\vspace{5mm}
\facilities{VLA}

\clearpage

%% This command is needed to show the entire author+affilation list when
%% the collaboration and author truncation commands are used.  It has to
%% go at the end of the manuscript.
%\allauthors

%% Include this line if you are using the \added, \replaced, \deleted
%% commands to see a summary list of all changes at the end of the article.
\listofchanges


\begin{thebibliography}{}

\bibitem[Alexander(2008)]{ale08} Alexander, R.\ 2008, MNRAS, 391, L64
\bibitem[Alexander et al.(2005)]{ale05} Alexander, R.~D., Clarke, C.~J., \& Pringle, J.~E.\ 2005, MNRAS, 358, 283
\bibitem[Alexander et al.(2014)]{ale14} Alexander, R., Pascucci, I., Andrews, S., Armitage, P., \& Cieza, L.\ 2014, in Protostars and Planets VI, ed. H. Beuther et al. (Tucson, AZ: Univ. Arizona Press), 475
\bibitem[Andrews et al.(2013)]{and13} Andrews, S.~M., Rosenfeld, K.~A., Kraus, A.~L, Wilner, D.~J.\ 2013, \apj, 771, 129
\bibitem[Anglada(1995)]{ang95} Anglada, G.\ 1995, RMxAC, 1, 67
\bibitem[Anglada et al.(2015)]{ang15} Anglada, G., Rodr\'{\i}guez, L.~F., \& Carrasco-Gonz\'alez, C.\ 2015, in Advancing Astrophysics with the Square Kilometre Array (AASKA14), 121
\bibitem[Avalos \& Lizano(2012)]{ava12} Avalos, M., Lizano, S.\ 2012, \apj, 751, 63
\bibitem[Beckwith \& Sargent(1991)]{bec91} Beckwith, S.~V.~W., Sargent, A.~I.\ 1991, \apj, 381, 250 
\bibitem[Cabrit(2007)]{cab07} Cabrit, S.\ 2007, in IAU Symp. 243, Star-Disk Interaction in Young Stars, ed. J. Bouvier \& I. Appenzeller (Cambridge: Cambridge Univ. Press), 203
\bibitem[Calvet et al.(2005)]{cal05} Calvet, N., D'Alessio, P., Watson, D.~M., et al.\ 2005, \apj, 630, L185
\bibitem[Carrasco-Gonz\'alez et al.(2012)]{car12} Carrasco-Gonz\'alez, C., Galv\'an-Madrid, R., Anglada, G., et al.\ 2012, \apjl, 752, L29
\bibitem[Clarke et al.(2001)]{cla01} Clarke, C. J., Gendrin, A., \& Sotomayor, M. \ 2001, MNRAS, 328, 485
\bibitem[Dutrey et al.(1998)]{dut98} Dutrey, A., Guilloteau, S., Prato, L., et al.\ 1998, A\&A, 338, L63
\bibitem[Ercolano \& Owen(2010)]{erc10} Ercolano, B., Owen, J.~E.\ 2010, MNRAS, 406, 1553
\bibitem[Espaillat et al.(2010)]{esp10} Espaillat, C., D'Alessio, P., Hern\'andez, J., et al.\ 2010, 717, 441
\bibitem[Espaillat et al.(2013)]{esp13} Espaillat, C., Ingleby, L., Furlan, E., et al.\ 2013, \apj, 762, 62
\bibitem[Font et al.(2004)]{fon04} Font, A., McCarthy, I., Johnstone, D., Ballantyne, D.~R.\ 2004, \apj, 607, 890
\bibitem[Galv\'an-Madrid et al.(2014)]{gal14} Galv\'an-Madrid, R., Liu, H.~B., Manara, C.~F.\ 2014, A\&A, 570, L9 
\bibitem[Gorti et al.(2009)]{gor09} Gorti, U., Dullemond, C. P., \& Hollenbach, D.\ 2009, \apj, 705, 1237
\bibitem[Gorti \& Hollenbach(2009)]{gor09b} Gorti, U., \& Hollenbach, D.\ 2009, \apj, 690, 1539
\bibitem[G\"udel et al.(2010)]{gud10} G\"udel, M., Lahuis, F., Briggs, K.~R., Carr, J., et al.\ 2010, A\&A, 519, 113
\bibitem[Hartigan et al.(1995)]{har95} Hartigan, P., Edwards, S., \& Ghandour, L.\ 1995, \apj, 452, 736
\bibitem[Hollenbach \& Gorti(2009)]{hol09} Hollenbach, D., Gorti, U.\ 2009. \apj, 703, 1203
\bibitem[Hughes et al.(2009)]{hug09} Hughes, A.~M., Andrews, S.~M., Espaillat, C., et al.\ 2009, \apj, 698, 131
\bibitem[Ingleby et al.(2015)]{ing15} Ingleby, L., Espaillat, C., Calvet, N., et al.\ 2015, \apj, 805, 149
\bibitem[Kenyon \& Hartmann(1995)]{ken95} Kenyon, S.~J., Hartmann, L.\ 1995, \apjs, 101, 117
\bibitem[Kitamura et al.(2002)]{kit02} Kitamura, Y., Momose, M., Yokogawa, S., et al.\ 2002, \apj, 581, 357
\bibitem[Looney et al.(2000)]{loo00} Looney, L.~W., Mundy, G., \& Welch, W.~J.\ 2000, \apj, 529, 477
\bibitem[Lugo et al.(2004)]{lug04} Lugo, J., Lizano, S., \& Garay, G.\ 2004, \apj, 614, 807
\bibitem[Najita et al.(2009)]{naj09} Najita, J.~R., Doppmann, G.~W., Bitner, M.~A., Richter, M.~J., et al.\ 2009, \apj, 697, 957
\bibitem[Owen et al.(2011)]{owe11} Owen, J.~E., Ercolano, B., Clarke, C.~J.\ 2011, MNRAS, 412, 13
\bibitem[Owen et al.(2010)]{owe10} Owen, J.~E., Ercolano, B., Clarke, C.~J., \& Alexander, R.~D.\ 2010, MNRAS, 401, 1415
\bibitem[Owen et al.(2013)]{owe13} Owen, J.~E., Scaife, A.~M.~M., Ercolano, B.\ 2013, MNRAS, 434, 3378
\bibitem[Pascucci et al.(2012)]{pas12} Pascucci, I., Gorti, U., \& Hollenbach, D.\ 2012 \apjl, 751, L42
\bibitem[Pascucci et al.(2014)]{pas14} Pascucci, I., Ricci, L., Gorti, U., et al.\ 2014 \apj, 795, 1
\bibitem[Pascucci \& Sterzik(2009)]{pas09} Pascucci, I., \& Sterzik, M.\ 2009 \apj, 702, 724
\bibitem[Rau \& Cornwell(2011)]{rau11} Rau, U., Cornwell, T.~J.\ 2011, A\&A, 532, 71
\bibitem[Rodmann et al.(2006)]{rod06} Rodmann, J., Henning, T., Chandler, C.~J., et al.\ 2006, A\&A, 446, 211
\bibitem[Rodr\'{\i}guez et al.(2008)]{rod08} Rodr\'{\i}guez, L.~F., Torrelles, J.~M., Anglada, G., Reipurth, B.\ 2014, AJ, 136, 1852
\bibitem[Rodr\'{\i}guez et al.(2014)]{rod14} Rodr\'{\i}guez, L.~F., Zapata, L.~A., Dzib, S.~A., et al. 2014, ApJL, 793, L21
\bibitem[Sacco et al.(2012)]{sac12} Sacco, G.~G., Flaccomio, E., Pascucci, I., et al.\ 2012, \apj, 747, 142
\bibitem[Salyk et al.(2013)]{sal13} Salyk, C., Herczeg, G.~J., Brown, J.~M., et al.\ 2013, \apj, 769, 21
\bibitem[Weintraub et al.(1989)]{wei89} Weintraub, D.~A., Sandell, G., Duncan, W.~D.\ 1989, \apj, 340, L69
\bibitem[Williams \& Cieza(2011)]{wil11} Williams, J.~P., Cieza, L.~A.\ 2011, A\&A, 49, 67

\end{thebibliography}
\end{document}